\documentclass[12pt]{article}
\usepackage{epsfig,amsfonts,amssymb}
\usepackage{hyperref}
\pdfoutput=1

\usepackage{cite}
\usepackage{cite}

\usepackage{comment}

\def\ZZZ{{\hbox{ Z\kern-1.6mm Z}}}
\def\RRR{{\hbox{ R\kern-2.4mm R}}}
\def\CCC{{\hbox{ C\kern-2.0mm C}}}
\def\zzz{{\hbox{z\kern-1mm z}}}

\def\ZZZ{\mathbb{Z}}
\def\RRR{\mathbb{R}}

\newcommand{\qeq}{{\hbox{=\kern-2.3mm ? \kern.5mm }}}
\renewcommand{\qeq}{=}

\newcommand{\eps}{\epsilon}

\newcommand{\ve}{\varepsilon}

\newcommand{\GG}{{\cal G}}

\newcommand{\OO}{{\cal O}}

\newcommand{\EE}{{\cal E}}
\newcommand{\LL}{{\cal L}}

\newcommand{\wt}{\widetilde}
\newcommand{\wh}{\widehat}

\newcommand{\NN}{{\cal N}}

\newcommand{\SSS}{{\cal S}}

\newcommand{\be}{\begin{equation}}
\newcommand{\ee}{\end{equation}}
\newcommand{\ben}{\begin{eqnarray}\displaystyle}
\newcommand{\een}{\end{eqnarray}}

\newcommand{\refb}[1]{(\ref{#1})}
\newcommand{\p}{\partial}
\newcommand{\sectiono}[1]{\section{#1}\setcounter{equation}{0}}

\def\one{{\hbox{ 1\kern-.8mm l}}}
\def\zero{{\hbox{ 0\kern-1.5mm 0}}}

\newcommand{\bea}[1]{\begin{eqnarray}\label{#1} }
\newcommand{\eea}{\end{eqnarray}}

\newcommand{\eqref}{\refb}

\newcommand{\hh}{\hat h}

\newcommand{\nablax}{\p}

\usepackage{bm}
\usepackage[table]{xcolor}

\begin{document}

\baselineskip 24pt

\begin{center}

%{\Large \bf How to Study Isolated Exotic Branes}
{\Large \bf Gravitational Path Integral and String Theory}

\end{center}

\vskip .6cm
\medskip

\vspace*{4.0ex}

\baselineskip=18pt

\centerline{\large \rm Ashoke Sen}

\vspace*{4.0ex}

\centerline{\large \it International Centre for Theoretical Sciences - TIFR 
}
\centerline{\large \it  Bengaluru - 560089, India}

%\centerline{\large \it ~$^c$Homi Bhabha National Institute}
%\centerline{\large \it Training School Complex, Anushakti Nagar,
%    Mumbai 400085, India}

\vspace*{1.0ex}
\centerline{\small E-mail:  ashoke.sen@icts.res.in}

\vspace*{5.0ex}

\centerline{\bf Abstract} \bigskip

Gravitational path integral has had many applications in recent years.
Using the example of supersymmetric index, which provides one of the most controlled
tests of gravitational path integral, we argue that even in this controlled case, naive
gravitational path integral cannot produce consistent results beyond low orders. However
one can go further using stringy inputs, e.g. the choice of UV cut-off and the choice
of correct path integration variable, without having to do
full fledged string calculation.  
We also discuss to what extent we are led to string scale UV cut-off even if
we begin with a supergravity theory without an a priori bias towards string theory.
This is a review of mostly
published results, but we clarify several
details.

\vfill \eject

\tableofcontents

\sectiono{Introduction}

Gravitational path integral\cite{Gibbons:1976ue} 
has had many important applications in recent 
years (see, for example, \cite{1903.11115,1911.12333,2003.02860,2212.02447}).
Most of these have been semi-quantitative leading order results.
The goal of this article is to review at what level the path integral stops making sense 
and one needs to use information from ultraviolet completion.
In string theory this involves using inputs from string theory.

The central object in our analysis will be the supersymmetric 
index\cite{9611205,9708062}
that counts, with sign,
the number of microstates of a supersymmetric black hole. Even if one is not
interested in such index, it provides a useful testing ground for gravitational path
integral since it can be computed from path integral
in a systematic fashion\cite{1810.11442,2107.09062,2306.07322}
and at the same time one knows the exact results for the
index from microscopic counting in some cases. Attempts have been made to
evaluate this gravitational path integral using supersymmetric 
localization\cite{0905.2686,1009.3226,1012.0265,
1111.1161,1404.0033,1806.04479,2209.13602}
without invoking stringy UV completion of the theory. Technical criticism of this approach
has been discussed elsewhere\cite{2302.13490}, 
but even without going into these technicalities, one
can see (as will be reviewed later) that the one loop contribution to the index is
sensitive to the UV cut-off and one gets the wrong answer even at this level unless
one uses the string scale cut-off.\footnote{Since
\cite{2209.13602} used Planck scale cut-off in their analysis, one could ask how they
got the correct answer.
One immediate issue that one can point out in their
analysis involves integration over gauge field zero modes.  
If a gauge field has an effective (moduli dependent) coupling $g_{\rm gauge}$, 
then integration over each gauge field zero mode will produce a factor of 
$g_{\rm gauge}^{-1}$ besides
appropriate power of the cut-off. This was not taken into account in the localization
analysis.
It is precisely these types of factors that force us to choose string scale as
the UV cut-off. As we shall see, unlike in the near horizon $AdS_2\times S^2$
set up of \cite{2209.13602} where the gauge fields have zero modes, in the
asymptotically flat geometry we do not have gauge field zero modes. But the
string coupling dependent contribution arises from the 2-form zero modes and 
during the process of  changing ensembles from grand canonical to canonical.}
If even for a controlled quantity like the index one needs
stringy input, it seems unlikely that for more general quantities one can achieve
more than semi-quantitative success based on gravitational path integral without
UV completion. A drastic scenario of this type would be 
that in the path integral one might identify a 
saddle by finding classical solution, but the one loop determinant, after taking into
account the UV completion, may vanish making the whole contribution disappear.

Our goal in this review will be to analyze the one loop correction to the 
index of supersymmetric black holes in string theory. By adjusting the charges and the
asymptotic values of the moduli one can ensure that the black hole saddle that 
contributes to the index has large size and small string coupling. In that case one can
compute corrections to the entropy involving logarithms of size of the black hole and
logarithm of string coupling from low energy physics and the knowledge of UV cut-off,
without having to do a full fledged stringy calculation.
We show that if the UV cut-off is chosen to be the string scale then the results are in perfect
agreement with the microscopic counting results as well as general properties of the index
expected from supersymmetry. Conversely, 
if we set out to compute the logarithmic correction to
the index in an extended supergravity theory without any {\it a priori} bias 
towards string theory, we are led to the choice of string scale as the UV cut-off in order to
satisfy the desired property of the index like its robustness under variation of the asymptotic
values of the moduli fields.

In principle one should be able to go beyond this analysis by doing a full one loop string
computation in the black hole background. This will produce, besides the logarithmic terms,
the complete correction to
the index that
remains finite in the limit of large size and small string coupling. To the best of our knowledge,
such a computation has not been done. What has been done is to compute one loop
correction to some terms in the effective action in flat 
background\cite{9610237,9708062} and then use those
corrections to compute corrections to the black hole entropy / index and compare with
known microscopic results\cite{9607026,0412287,0510147,0607155,0609109}. 
But this leaves open the possibility that other terms
in the action can also contribute at this order. Some general arguments have been made
as to why they may not contribute\cite{1306.3796}, but there is no rigorous proof.

The review is organized as follows. For definiteness we restrict our analysis to four
space-time dimensions although in principle this can be generalized to higher dimensions
as well. 
Also, in our analysis we assume, for convenience, that the asymptotic values of the
moduli fields are set equal to their attractor values.
In section \ref{s2} we discuss charge quantization,
emphasizing that the data for the charge lattice are parameters of the theory, 
independent of the asymptotic values of the moduli fields. In a given theory the charge
lattice is fixed and no observer can change it, while the asymptotic values of
the moduli can be (effectively) changed by an observer. In section \ref{s3} we define the
index and its protection. Here we emphasize that the index should be independent of the
asymptotic values of the moduli but there is no reason to expect it to be independent of the
parameters labelling the charge lattice. In section \ref{s4} we describe the general
procedure for computing the index from gravitational path integral. 
In section \ref{s5} we explicitly evaluate the one loop contribution to the gravitational path
integral that is needed to compute the index, with special attention to the sensitivity
to the UV cut-off. In particular we take the UV cut-off length to be the string length times
$g_s^\gamma$ where $g_s$ is the string coupling and $\gamma$ is a constant. 
In section \ref{s6} we use these results to compute the index of two different types of
supersymmetric black holes in $\NN=2,4,6$ and $8$ supersymmetric string theories in
$D=4$, -- those carrying pure NSNS charges and those carrying pure RR charges.
In the first case the string coupling gets fixed in terms of the ratios of the
charges and we find that
the logarithmic correction to the logarithm of the index agrees with the result of microscopic
counting, known for $\NN=8$ theory, only when $\gamma=0$, i.e.\ when the UV
cut-off is taken to be the string scale. 
In the $\NN=4$ theory, where the microscopic results are also known, the path integral
results are insensitive to the UV cut-off and the results of path integral computation
agree with the results of microscopic computation.
For black holes carrying pure RR charges, the string coupling does not get
fixed in the black hole background and can be set to any constant value. Hence the
index should be independent of the string coupling. 
This can be tested in all the theories since this does not require knowing the microscopic
results.
We find that in all the theories, other than the $\NN=4$ theories, this
happens only if we choose $\gamma=0$, i.e.\ the UV cut-off is taken to be the string scale.
In the $\NN=4$ theory the result is insensitive to the choice of
UV cut-off.

While the analysis in section \ref{s6} indicates the need for using string scale UV cut-off,
it uses some inputs from string theory, {\it e.g.}  integration over the NSNS 2-form field 
instead of the axion, microscopic counting results in string 
theory in section \ref{s6.1} etc. In section \ref{s7} we start with supergravity theory
without any prejudice about the UV completion and attempt to set up a general set of
criteria that a UV completion of the theory should satisfy. In this process we also
develop a procedure for determining, within supergravity, which of the fields among
a dual pair we should integrate over in the path integral. This in particular justifies use
of the 2-form field as the integration variable within supergravity. Working 
within this framework, we find that for $\NN=6$ and 8 supergravities string scale 
automatically emerges as the UV cut-off without any prior input from string theory.
$\NN=4$ supergravity is agnostic to the UV cut-off, while for $\NN=2$ 
supergravity theory we need additional
input from string theory besides choosing the UV cut-off to be the string scale.

As mentioned before,
in our analysis we assume, for convenience, that the asymptotic values of the
moduli fields are set equal to their attractor values. It was argued in \cite{2605.30417} that in 
$\NN=4,8$ supergravity and in a
special class of $\NN=2$ supergravity known as the STU model, the moduli fields
remain constant throughout the solution when their
asymptotic values  are set equal to their attractor values. 
In appendix \ref{sa} we prove this for a general $\NN=2$ supergravity theory.
Constancy of the moduli fields is not necessary for our analysis as long as the
variations are of order unity, but having constant scalars makes the solution simpler.

\sectiono{Charge quantization and scalar moduli} \label{s2}

Typically extended supergravity theories in 3+1 dimensions contain a set of
massless scalars without any potential, known as moduli fields, and a set of
$U(1)$ 
gauge fields, collectively denoted by $A_\mu^{(i)}$, with gauge transformation
laws
\be \label{egauge}
\delta A_\mu^{(i)} =\p_\mu \Lambda^{(i)}\, .
\ee
We shall assume that the UV completion of the theory contains states charged under
these gauge fields. Otherwise there will be no charged black holes and the issues of the
kind we shall be addressing will be absent. A state carrying charges $\{Q^{(i)}\}$ under the
U(1) gauge fields will have a world line coupling of the form
\be\label{eworld}
\sum_i Q^{(i)} \int A_\mu^{(i)} dx^\mu\, .
\ee
This is manifestly invariant under the gauge transformations \refb{egauge}. For this it is
essential that $Q^{(i)}$'s remain constant along the world-line, and are not functions of the
world-line coordinates $x^\mu$. Typically the allowed values of $Q^{(i)}$ lie 
on a charge lattice and its magnetic dual charges, describing the total magnetic flux
through the celestial sphere, take values on the dual lattice. The lattice spacings are fixed in
a given theory and are genuine parameters of the theory. In particular the same
supergravity theory may have more than one UV completion with different charge lattice.
In a given theory, an observer within the theory cannot change the lattice spacing. In contrast
the asymptotic values of the moduli fields are parameters labelling the vacua of the
theory, but different values of the moduli fields  should be thought of as different states
of the same theory. In particular an observer can effectively change the 
asymptotic values of the moduli fields by creating a large region where the moduli
take the desired values and perform any experiment in that region to any 
desired accuracy\cite{2506.13876}.

Sometimes the supergravity theory under consideration may have global symmetries
that rotate the gauge fields and transform the moduli scalar fields leaving the action
invariant. Such a transformation would rotate the charge lattice, but will not generate a
new theory. The most famous example of this involves heterotic
string theory on $T^6\times R^{3,1}$
where charges take value on an even, self-dual Lorentzian lattice and 
all such lattices are related by $SO(6,22)$ transformations that can be undone by
a simultaneous rotation of the gauge fields 
and scalars\cite{Narain:1985jj,Narain:1986am}. 
Therefore these different
choices of lattices can be regarded as different asymptotic values of moduli fields.
But there are other possible deformations of the lattice, e.g. the ones that take
us outside the set of even, self-dual lattices, that may not be interpretable this way.
Of course most of them may not have UV completion, but determining this would
require more detailed analysis than what we shall be performing here. 
Even in the cases where the change in lattice can be regarded as
change in the asymptotic values of the moduli, we shall use the 
uniform convention that
the lattice is
fixed and the fields are variables. When we require
an observable like the index to be independent of the `parameters', it will always be
with respect to variation of the asymptotic values of the moduli and not with respect
to other parameters like the charge lattice, even though sometimes one can be traded in
for the other.

For practical  reasons sometimes it may be useful to
define a new set of gauge fields $\wt A^{(i)}_\mu$ via the relation
\be
A^{(i)}_\mu = \sum_j f^{(ij)}(\chi) \wt A^{(j)}_\mu\, ,
\ee
where $f^{(ij)}$ are functions of the moduli fields, collectively denoted as $\chi$.
Such a redefinition may be needed, {\it e.g.} if we want to canonically
normalize the kinetic term
of the gauge fields.
Then we can define 
\be
\wt Q^{(j)} = \sum_i Q^{(i)} f^{(ij)}(\chi)\, ,
\ee 
and write the world-line action \refb{eworld} as,
\be\label{eworldnew}
\sum_j  \int \wt Q^{(j)}  \wt A_\mu^{(j)} dx^\mu\, .
\ee
The new charges $\wt Q^{(j)}$ will now take values on a different lattice that depends
on the moduli fields $\chi$. In particular if the world-line passes through a background
where $\chi$ varies, the charges will not remain constant along the world-line. What remain
constant are the integers labelling the charges, and the values of $\wt Q^{(j)}$ will
change along the trajectory according to the way the lattice is deformed along the
trajectory. This can be traced to the fact that the new gauge fields $\wt A^{(j)}_\mu$ have
non-standard gauge transformation laws and hence the charges that couple to them are not
conserved by themselves. Even though this description may have practical utility, we
shall still parametrize the theory by the charge lattice on which the original
charges $Q^{(j)}$'s take value
and  the asymptotic values of the moduli fields.  In this description the charge lattice
is kept fixed in a given theory and the asymptotic values of the moduli
are allowed to vary since an observer can effectively change these values.

\sectiono{Supersymmetric index and its parameter dependence} \label{s3}

In extended supergravity theories, the supersymmetric index plays an important role
in the counting of (black hole) microstates. Let us suppose that the theory has a total number
of $4\NN$ supercharges and that we want to study states that preserve $4\NN -2n$ of these
supersymmetries. Then the number of supersymmetries that are broken by the state
is $2n$. In asymptotically flat 3+1 dimensional space-time, in which we shall work,
we define
the helicity supertrace\cite{9611205,9708062} to be
\be \label{ebndef}
B_n(q,\beta) = {(-1)^{n/2}\over n!}\, Tr_{q,\vec k=0}[(-1)^F e^{-\beta H} (2h)^n]\, ,
\ee
where $Tr_{q,\vec k=0}$ denotes that the trace is taken over all states at rest, 
carrying fixed electric and magnetic U(1) charges collectively denoted as
$q$, and $h$ is the third component of the angular momentum in the rest frame. 
One can show that this receives contribution only from states that break $2n$ or
less supersymmetries. Furthermore for generic charges, states that break strictly
less than $2n$ supersymmetries do not exist and  $B_n$
gets contribution only from states that break exactly
$2n$ supersymmetries. Our focus in this article will be on 
supersymmetric black holes with finite area event horizon. These break four
supersymmetries and hence we need $2n =4\NN-4$.
In particular, we have $2n=28$ for $\mathcal N=8$ supersymmetric theories, 
$2n=20$ for $\mathcal N=6$ supersymmetric theories, 
$2n=12$ for $\mathcal N=4$ supersymmetric theories
and $2n=4$ for $\mathcal N=2$ supersymmetric theories. Since the states
contributing to \refb{ebndef} all have the same mass $M(q)$ given by the BPS
mass formula, 
we get\footnote{It was suggested 
in \cite{1406.2360,1501.01643,1808.05606} that at finite temperature, the index
$B_{n}$ of the type we are considering can receive contribution not only from 
single centered black hole and their bound states but also
scattering states involving multi-centered black holes. 
The latter contribution is temperature dependent since their total energy differs
from that of BPS bound states.  However such contributions
are typically exponentially sub-dominant for large black holes that we shall be
analyzing and hence we can ignore this effect.}
\be\label{ebqexp}
B_{n}(q,\beta) = e^{S_{BPS}(q) - \beta \, M(q)}\, ,
\ee
where $M(q)$ is the mass of the BPS state of charge $q$, giving the eigenvalue of 
$H$ at $\vec k=0$ and $e^{S_{BPS}}$ denotes the number of supermultiplets
with zero momentum and fixed charge $q$,
breaking $2n$ supersymmetries, counted with the weight,
\be
{(-1)^{n/2}\over n!}\, Tr_{\rm supermultiplet} [(-1)^F (2h)^n] = \pm 1\, .
\ee
Here + sign is chosen if the average $J_3$ eigenvalue of all the states 
in the supermultiplet is an integer and the $-$ sign is chosen if the average $J_3$ eigenvalue of all the states 
in the supermultiplet is integer + half.
Therefore we can compute $S_{BPS}$ using the relation
\be
S_{BPS} = \ln B_{n}(q,\beta) + \beta \, M(q)\, .
\ee
At the leading order $S_{BPS}$ can be related to the Bekenstein-Hawking entropy
of a supersymmetric black hole.

Since $S_{BPS}$ is a protected quantity, it is expected to be unchanged under
continuous variation of the parameters of the theory. However one should be
careful about what one means by parameters of the theory. For example from the
point of view of supergravity, the  lattice spacings of the charge lattice are
parameters of the theory, but $S_{BPS}$ is not invariant under such a change.
Indeed, if we change the lattice spacing keeping fixed the integers associated
with the charges, the entropy of the corresponding black hole will change and
$S_{BPS}$ will also change. This is not a contradiction since no observer within the
theory can change the lattice spacing. On the other hand, since an observer can
change the effective asymptotic values of the moduli, $S_{BPS}$ should be independent
of the asymptotic values of the moduli fields. In particular we could set up a background
with slowly varying moduli and let the supersymmetric black hole roll adiabatically
in this background. The number of states of the black hole should not change during this
process.

\sectiono{Index from gravitational path integral: General setup} \label{s4}

The starting point for the index computation is the euclidean path integral over all the
fields in the theory with the
following boundary condition on the massless fields:
\begin{enumerate}
\item The period of the Euclidean time circle is set equal to $\beta$.
\item The magnetic flux associated to various $U(1)$ fields is set equal to some
fixed magnetic charge vector $P$.
\item $\ointop A^{(i)}_\mu dx^\mu$, with the integration contour taken to be along the
asymptotic time circle, is set equal to $\beta\mu_i$. At present we take $\mu_i$ to be
an arbitrary complex number but the black hole saddle that will provide dominant
contribution to the path integral will have imaginary $\mu_i$.
\item The shift in the azimuthal angle $\phi$ as we go once around the Euclidean
time circle is taken to be equal to $i\omega\beta$. Eventually we shall take $\omega$
to be imaginary so that the shift is real.
\item The path integral is weighted by the usual factor of $e^{-I_E}$ where $I_E$ is the
Euclidean action. Besides this we also insert a factor of $(-1)^{n/2}
(2J_3)^n/n!$, where $J_3$ is the
third component of the angular momentum, expressed as a functional of the fields using
the Noether procedure.
\end{enumerate}
We shall denote by $Z_n$ the result of this path integral. According to Gibbons and
Hawking\cite{Gibbons:1976ue} this path integral will compute
\be
Z_n(\beta, P, \mu,\omega) = (-1)^{n/2}\, {1\over n!} \,
Tr_P[e^{-\beta E -\beta\mu.Q - \beta\omega h} (2h)^n]\, ,
\ee
where $Tr_P$ denotes that the trace is taken over states carrying fixed magnetic charges
$P$, $E$ denotes the energy of the state, $Q$ is the electric charge vector carried by 
the state and $h$ is the $J_3$ eigenvalue carried by the state. 
As long as the gauge fields $A^{(i)}$'s are normalized so that they have the standard
gauge transformation laws \refb{egauge}, 
$Q$ and $P$ are normalized so that in a given
theory they take values on a fixed lattice independent of the moduli, i.e.\ they are the
analogs of the charges $Q^{(i)}$ that appeared in \refb{eworld} and their dual. 
In order to make its
connection with $S_{BPS}$ explicit, we shall set $\omega = -2\pi i/\beta$ and
make explicit the sum over the electric
charges $Q$ and the integration over momenta $\vec k$ in the 
trace:\footnote{We could also set $\omega=-2\pi i/\beta$, but eventually we shall
integrate over the rotational zero modes that will include this value.}
\be
Z_n(\beta, P, \mu,\omega=2\pi i/\beta) =(-1)^{n/2}\,  {1
\over n!}\, \sum_Q \int L^3\, {d^3 k\over (2\pi)^3} \, 
Tr_{P,Q, \vec k} [e^{-\beta E(\vec k) -\beta\mu.Q + 2\pi i h} (2h)^n]\, ,
\ee
where we have implicitly assumed that the system is enclosed in a box of
linear size $L$ so that the density of states is given by $L^3 {d^3 k / (2\pi)^3}$.
Since $e^{2\pi i h}=(-1)^F$, the trace gets contribution only from BPS states
for which the mass is a given function $M(Q,P)$ of the charges. Furthermore due to
Lorentz invariance, the
number of quantum states is independent of the momentum $\vec k$. 
We now use the result
\be \label{eEk}
E(\vec k) = M(Q,P) + {\vec k^2\over 2 M(Q,P)} +\OO(\vec k^4)\, ,
\ee
to carry out the integration over $\vec k$
and get
\ben \label{eznb}
Z_n(\beta, P, \mu,\omega &=& 2\pi i/\beta) = (-1)^{n/2}{1
\over n!} {L^3\over (2\pi)^{3/2}} \{M(Q,P)/\beta\}^{3/2} \nonumber \\ && \hskip 1in
\sum_Q 
Tr_{P,Q, \vec k=0} [e^{-\beta M(Q,P)-\beta\mu.Q }(-1)^F (2h)^n]\, .
\een
The $\OO(\vec k^4)$ terms in \refb{eEk} produces terms suppressed by powers of $1/M$ in
\refb{eznb} that we have ignored.
Using \refb{ebndef}, \refb{ebqexp}, this can be expressed 
as\cite{1810.11442,2107.09062,2306.07322}
\be
Z_n(\beta, P, \mu,\omega=2\pi i/\beta)  = 
{L^3\over (2\pi)^{3/2}} 
\sum_Q  \{M(Q,P)/\beta\}^{3/2} e^{S_{BPS}(Q,P) -\beta M(Q,P)-\nu.Q}, \quad
\nu_i \equiv \beta \mu_i\, .
\ee
One can invert this relation to write:
\be
e^{S_{BPS}(Q,P)} \sim L^{-3} (\beta / M(Q,P))^{3/2} \, e^{\beta M(Q,P)} \,
\int d^{n_v}\nu \, e^{\nu.Q +\ln 
Z_n(\beta, P, \mu,\omega=2\pi i/\beta)} \, ,
\ee
where it is understood that the $\nu$ integration runs along the imaginary axis and
has a fixed range independent of the moduli, 
determined by the charge lattice. $n_v$ is the number of $U(1)$ gauge
fields and $\sim$ denotes equality up to a numerical
constant of proportionality that has no dependence on the values of the moduli fields,
but depends, for example, on the volume of the unit cell of the charge lattice.

One can now evaluate the integral over $\nu_i$'s and the path integral contribution to
$Z_n$ by saddle point method. 
First consider the path integral over $\nu_i$. 
We write the equation for the saddle point as
\be
Q_i = -{\p\over \p\nu_i} \ln Z_n\, .
\ee
After performing the $\nu_i$ integral, we get, to first subleading order,
\be \label{e4.8xy}
e^{S_{BPS}(Q,P)} \sim L^{-3} (\beta / M(Q,P))^{3/2} 
e^{\beta M(Q,P) +\nu.Q + \ln 
Z_n}
\left(\det {\p^2 \ln Z_n\over \p \nu_i\p\nu_j}\right)^{-1/2}\, .
\ee
In the path integral computation of $Z_n$, 
the dominant contribution
comes from a Euclidean rotating black hole saddle point for given values of
$\beta$, $\omega$ and $\nu_i$'s, satisfying $\beta\omega=2\pi i$.
We shall denote the classical contribution to $\ln Z_n$, given by the negative of the
Euclidean action of the black hole solution,
by $\ln Z_n^{\rm cl}$ and the first subleading correction by $\Delta\ln Z_n$. Then while
evaluating the last term in \refb{e4.8xy}, it is enough to work with 
$\ln Z_n^{\rm cl}$. Hence we get
\ben \label{e4.9xy}
S_{BPS}(Q,P) &\simeq& \beta M(Q,P) +\nu.Q + \ln 
Z_n^{\rm cl}+  \Delta \ln Z_n \nonumber \\
&-& 3\, \ln L +{3\over 2} \ln {\beta\over M(Q,P)}
-{1\over 2} \ln \det {\p^2 \ln Z^{\rm cl}_n\over \p \nu_i\p\nu_j}\, .
\een

For simplicity, we shall take the asymptotic values of the scalars to be their
attractor values.
In that case the scalars remain constant throughout the solution and the
metric becomes that of the Euclidean Kerr-Newman solution with 
$\beta\omega=2\pi i$:\footnote{This result was found for the $\NN=2$ 
supersymmetric STU model and $\NN=4,8$ supergravity in \cite{2605.30417}. 
We show in
appendix \ref{sa} that this is a general result in $\NN=2$ supergravity.
We note, however, that having constant scalars 
is not necessary, -- our analysis will remain valid as
long as the scalars do not vary drastically in the solution.}
\be\label{eprestring}
g_{\mu\nu} = a^2 g^{(0)}_{\mu\nu}\, ,
\ee
where 
\ben\label{eg0}
g^{(0)}_{\mu\nu} dx^\mu dx^\nu &=&  {  r^2 - b^2 \cos^2\theta \over (r+1)^2 - b^2 \cos^2\theta}\, 
 (d\tau+\wt\omega_\phi d\phi) ^2 \nonumber \\
&& \hskip -.5in+\  {(r+1)^2 - b^2 \cos^2\theta \over r^2 - b^2 \cos^2\theta} \left[
{{r^2-b^2\cos^2\theta} \over {r^2-b^2}}dr^2+(r^2-b^2\cos^2\theta)d\theta^2
+(r^2-b^2)\sin^2\theta d\phi^2\right]\, , \nonumber \\
\wt\omega_\phi &=& b\, \sin^2\theta \, {2\, r +1\over r^2 - b^2\cos^2\theta}\, .
\een
$a$ in \refb{eprestring}  and $b$ in \refb{eg0} are
determined in terms of the charges carried by the black hole and the inverse temperature
$\beta$.
If we scale 
all the charges and $\beta$ by
$\lambda$,  $a$ scales by $\lambda$ and $b$ remains fixed.
The final result for the index is expected to be 
independent of $\beta$, but we shall take $\beta$ to be of order $a$. 
This corresponds to taking $b$ to be of order unity.
In this case 
the black hole saddle has a single large scale $a$ that
appears as an overall scale factor in the metric.

So far we have not specified which metric $g_{\mu\nu}$ refers to.
Since for our solution the scalars are constant throughout the solution, the dilaton
$\phi$ takes the same value everywhere in the solution and we denote by $g_s$ the
background value of $e^\phi$. Then the string frame metric $G_{\mu\nu}$
and the Einstein
frame metric $G^E_{\mu\nu}$ differ by a constant factor $g_s^2$. We write
\be\label{estring}
G_{\mu\nu} = \lambda_S^2 g^{(0)}_{\mu\nu}\, ,
\ee
for some
fixed large number $\lambda_S$ determined by the
charges. The Einstein frame
metric is then given by 
\be
G^E_{\mu\nu}=g_s^{-2} \lambda_S^2 g^{(0)}_{\mu\nu}\, .
\ee

Although the final result for $S_{BPS}$ will be independent of which metric we use to
measure various length and mass scales, it will be useful to fix a convention. We shall
use the convention that we use the string frame metric to measure all scales, i.e.
$\beta,M,L$ etc. are all measured in the string frame. In that case $L^3$ is the volume
of space-time measured in the string metric and $\beta$ is the period of the Euclidean
time circle and $M$ is the mass of the black hole measured in the string frame.
Since in this frame the Newton's constant $G_N$ is of order $g_s^2$, the
Schwarzschild radius of the black hole is of order $g_s^2M$. On the other hand for 
$b\sim 1$, we see from \refb{eg0} and \refb{estring} that the period $\beta$
of the Euclidean time circle and the Schwarzschild radius of the black hole are both
of order $\lambda_S$. Hence we have
\be \label{escal}
g_s^2 M\sim \lambda_S, \qquad \beta\sim \lambda_S \, .
\ee

For such a background we can estimate the contribution to 
$\ln \det {\p^2 \ln Z^{\rm cl}_n\over \p \nu_i\p\nu_j}$ as follows. Since the euclidean
time coordinate $\tau$ has period of order unity, it follows from the definition of 
$\beta\mu_i$ given at the beginning of this section that asymptotically
$A_\tau^{(i)}\sim \beta\mu_i=\nu_i$. On the other hand $A_\tau^{(i)}$ vanishes at the
horizon. Hence the gauge field strength is of order $\nu_i$. Since the NSNS sector
kinetic terms are multiplied by factors of $e^{-2\phi}$, the $\nu_i$ dependent
part of the euclidean action is of order $g_s^{-2} \nu_i\nu_j$. The dependence on the
size parameter $\lambda_S$ drops out due to scale invariance of the Maxwell action.
On the other hand
the RR sector gauge field kinetic terms do not have the $e^{-2\phi}$ factor and for
them the action contains a term of order $\nu_i\nu_j$. This gives
\be
\det {\p^2 \ln Z^{\rm cl}_n\over \p \nu_i\p\nu_j} \sim g_s^{-2 n_{NS}}\, ,
\ee
where $n_{NS}$ is the number of NS sector gauge fields. Hence
\refb{e4.9xy} takes the form:
\be \label{efinfor}
S_{BPS}(Q,P) \simeq S_{cl}  - 
3\, \ln L +3\ln g_s+  \Delta \ln Z_n
+ n_{NS} \ln g_s\, ,
\ee
where we used \refb{escal} to evaluate $\beta/M$, and
\be
S_{cl} = \beta M(Q,P) +\nu.Q + \ln 
Z_n^{\rm cl}\, ,
\ee
is the classical contribution to $S_{BPS}$. $S_{cl}$ is known to be independent of
$\beta$ and the asymptotic values of the moduli fields\cite{2610.02306}.
In the next section 
we shall discuss the computation of $\Delta\ln Z_n$. 

\sectiono{UV sensitive contribution to the index} \label{s5}

We now discuss the computation of $\Delta\ln Z_n$. 
This comes from integration over
the fluctuating fields around the classical saddle. 
For the first subleading contribution it is enough to expand the action in powers of the
fluctuating fields and keep up to quadratic terms in the fluctuations.

\subsection{Non-zero mode contribution}

First we consider the contribution from the bosonic fields. Let $K_b$ denote the kinetic term
for all the bosonic fields including bosonic ghosts
and $\{\lambda_b\}$ be the set of all the eigenvalues of $K_b$. 
Then up to
an overall constant normalization, the
result of integration over the modes with non-zero  eigenvalues is given by
\be
{\prod_b}' \lambda_b^{-1/2}\, ,
\ee
where the prime on the product means that we take the product over the non-zero
eigenvalues only. Similarly, if $\{\lambda_f\}$ denote collectively the eigenvalues of
the fermionic kinetic operator, then the result of integration over the fermionic modes 
with non-zero eigenvalues is
\be
{\prod_f}' \lambda_f^{1/2}\, .
\ee
Therefore, the contribution to $\Delta\ln Z_n$ from the non-zero modes is given by
\be \label{enonzero}
- {1\over 2} {\sum_b}' \ln \lambda_b + {1\over 2} {\sum_f}' \ln \lambda_f\, .
\ee
Typically these sums suffer from ultraviolet divergences since $\lambda_b$ and $\lambda_f$
are unbounded from above. 
A related issue is that $\lambda_b$ and $\lambda_f$ are dimensionful quantities,
with $\lambda_b$ having dimension of length$^{-2}$ and $\lambda_f$ having dimension
of length$^{-1}$, and hence computing logarithms of these terms requires a
scale. Both problems can be addressed by regulating the sum using heat
kernel regularization (see \cite{0306138} for a review),
which also resonates with
string theory. We use the identity:
\be\label{eheat}
\int_\eps^\infty s^{-1} ds e^{-\lambda s} = \cases{-\ln(\lambda\eps) \hbox{ + constant \quad
for
$\lambda\eps << 1$}\cr 0 \quad \hbox{for $\lambda\eps>>1$} }
\ee
If $\lambda$ is one of the $\lambda_b$'s then $\eps$ has the dimension of length$^2$.
Now if we replace $\ln\lambda_b$ by the left hand side of \refb{eheat}, then the sum
effectively gets cut-off at $\lambda_b\sim \eps^{-1}$, showing that $\sqrt\eps$ provides
an ultraviolet length cut-off. We call this $ l_{UV}$. A similar analysis follows for
$\lambda_f$ except that since $\lambda_f$ has dimension of length$^{-1}$, we
work with $\lambda_f^2$. This replaces \refb{enonzero} by
\be \label{enonzerox}
{1\over 2} \int_{ l_{UV}^2}^\infty s^{-1} ds {\sum_b}' e^{-\lambda_b s}  
- {1\over 4} \int_{ l_{UV}^2}^\infty s^{-1} ds {\sum_f}' e^{-\lambda_f^2 s}  
\simeq
- {1\over 2} {\sum_{b\atop \lambda_b< l_{UV}^{-2}}}' \ln (\lambda_b l_{UV}^2) + {1\over 2} {\sum_{f
\atop \lambda_f <  l_{UV}^{-1}}}' \ln (\lambda_f l_{UV})\, .
\ee

Since $\lambda_b l_{UV}^2$ and $\lambda_f l_{UV}$ are dimensionless, it does not matter
what metric we use to compute these as long as we use the same metric to compute
the eigenvalues and $l_{UV}$. 
For definiteness let us use the string metric. In that case it follows from
\refb{estring} that 
\be \label{eonex}
\lambda_b\sim \lambda_S^{-2}, \qquad \lambda_f\sim \lambda_S^{-1}\, .
\ee
In string theory, $l_{UV}\sim 1$ in string units (and $s$ becomes the moduli parameter
$\tau_2$ of the world-sheet torus) but since our goal is to demonstrate sensitivity
to the UV cut-off, we shall take
\be \label{etwox}
l_{UV} = g_s^\gamma\, ,
\ee
for some constant $\gamma$. If we denote by $N^b_{tot}$ the total number of bosonic
modes, by $N^f_{tot}$  the total number of fermionic
modes, by $N^b_z$ the total number of bosonic
zero modes and by $N^f_z$  the total number of fermionic
zero modes, then we get, from \refb{enonzerox}, \refb{eonex} and
\refb{etwox} that the non-zero mode contribution
to $\Delta\ln Z_n$ is of the form
\be\label{endiff}
\Delta_{nz}\ln Z_n = \left(N^b_{tot}-N^b_z -{1\over 2} N^f_{tot} 
+{1\over 2} N^f_z\right) \ln (\lambda_S g_s^{-\gamma})+\OO(1)\, .
\ee
Computation of the $\OO(1)$ terms will require full use of string theory, which
we shall not attempt to do here.

As it stands, \refb{endiff} is ambiguous since the definitions of $N^b_{tot}$ and
$N^f_{tot}$ depend on how exactly we cut off the sums on the right hand side of
\refb{enonzerox}. However it turns out that with the help of the left hand side of 
\refb{enonzerox} one can
give a precise definition of $N^b_{tot}$ and
$N^f_{tot}$ using a method known as heat kernel expansion.\footnote{In general
$N^b_{tot}$ and
$N^f_{tot}$ computed this way are fractional.}
We shall not
review this here, but only note that for $\NN\ge 2$ supergravity theories the result
is proportional to the Euler number of the black hole space-time with the coefficient of
proportionality determined by the spectrum of 
supermultiplets\cite{1505.01156,1905.13058}. 
We give the
final results here\cite{2306.07322,2605.30417,2606.17149}:\footnote{The results
in asymptotically flat geometry are computed using the known results in the
near horizon $AdS_2\times S^2$ 
geometry\cite{1005.3044,1106.0080,1108.3842,1109.0444} and then using the 
result\cite{1505.01156,1905.13058,2012.12227} 
that for
$\NN\ge 2$ supersymmetry $N^b_{tot}-{1\over 2} N^f_{tot}$ is proportional to the Euler
number of the Euclidean space time which takes the same value in the asymptotically
flat geometry and near horizon geometry.}
\be\label{entot}
N^b_{tot}-{1\over 2} N^f_{tot} = \cases{-10 \quad \hbox{for $\NN=8$}\cr
-6 \quad \hbox{for $\NN=6$}\cr -2 \quad \hbox{for $\NN=4$}\cr
{1\over 6}(11-n_V+n_H) \quad \hbox{for $\NN=2$}}\, ,
\ee
where $n_V$ and $n_H$ denote the number of vector multiplets and hypermultiplets
in the $\NN=2$ supersymmetric theory.

Before we describe the computation of $N^b_z$ and $N^f_z$, one comment is in order.
$N^b_{tot}$ and $N^f_{tot}$  are not invariant under duality transformation. For example
a scalar field can be dualized to a 2-form field in four dimensions but they give different
contribution to $N^b_{tot}$. However, it was shown in \cite{Duff:1980qv} 
that the difference comes
solely from the zero modes, -- the non-zero mode contribution is invariant under duality
transformation. \refb{entot} was computed in the description where the bosonic fields
are scalars, vectors and metric and the fermionic fields are spin 1/2 and spin 3/2. So as long
as we use the same description to calculate $N^b_z$ and $N^f_z$, \refb{endiff} will be free of
error.

In this description 
the bosonic zero modes include three translation modes and two rotation modes generated
by rotation about the $x$ and the $y$ axis if the black hole is spinning around the 
$z$ axis. Note that for generic rotation parameter for which $\phi$ gets shifted by a
generic angle as we go around the Euclidean time circle, these rotational modes will
not respect the periodicity condition. But for $\beta\omega=-2\pi i$ the shift of $\phi$ is
$2\pi$ under which the rotational modes are invariant. Therefore we have $N^b_z=5$.
On the other hand the fermionic zero modes come from the $2n$ broken supersymmetries
since each of them produces a goldstino mode. Thus we have 
\be\label{enz}
N^b_{z}-{1\over 2} N^f_{z} = 5 - n\, ,
\ee
where
$2n=4,12,20,28$ for $\NN=2,4,6,8$ supersymmetric theories.
Substituting \refb{entot} and \refb{enz} into \refb{endiff} we get the non-zero mode
contribution to $\Delta\ln Z_n$:
\be\label{endiffx}
\Delta_{nz} \ln Z_n = \cases{(-15+n) \ln (\lambda_S g_s^{-\gamma}) \quad \hbox{for $\NN=8$}\cr
(-11+n) \ln (\lambda_S g_s^{-\gamma}) \quad \hbox{for $\NN=6$}\cr 
(-7+n) \ln (\lambda_S g_s^{-\gamma}) \quad \hbox{for $\NN=4$}\cr
\left\{{1\over 6}(11-n_V+n_H)-5+n\right\} \ln (\lambda_S g_s^{-\gamma})
 \quad \hbox{for $\NN=2$}}\, .
\ee

\subsection{Zero mode contribution}

Next we turn to the contribution from the zero modes. For this we need to first
discuss an alternate but equivalent viewpoint on the UV cut-off. 
We have seen from \refb{enonzerox} that in the heat kernel regularization
 the contribution from each bosonic
non-zero mode is given by $-{1\over 2}\ln(\lambda_b l_{UV}^2)$ 
and the contribution from each fermionic
non-zero mode is given by ${1\over 2}\ln(\lambda_f l_{UV})$ where $l_{UV}$ is the
UV cut-off. A different way of stating this is to introduce a metric $\GG_{\mu\nu}$
in which $l_{UV}=1$ and then state that the contribution from each bosonic
non-zero mode is given by $-{1\over 2} \ln(\lambda_b)$ 
and the contribution from each fermionic
non-zero mode is given by ${1\over 2} \ln(\lambda_f)$ 
where $\lambda_b$ and $\lambda_f$ are
measured in the metric $\GG_{\mu\nu}$. If the UV cut-off is the string scale then
$\GG_{\mu\nu}$ is the string metric $G_{\mu\nu}$ but for more general choice of
cut-off \refb{etwox} we need 
\be \label{eggdef}
\GG_{\mu\nu}=e^{-2\gamma\phi}G_{\mu\nu} 
\, ,
\ee 
where $\phi$ is the dilaton field such that the expectation value of 
$e^\phi$ is $g_s$.

Had we been working in the harmonic gauge with diagonal kinetic operator, the above
prescription
would correspond to defining the result of the path integral as 
$(\det' [-\GG^{\mu\nu} D_\mu D_\nu])^{-1/2}$ and $(\det'[\wt\EE_a^\mu D_\mu])^{1/2}$
respectively,
where $\wt\EE_a^\mu$ is the inverse vierbein associated
with the metric $\GG_{\mu\nu}$. In general, in the presence of background
gauge fields the kinetic term may not have a simple
diagonal form but for counting powers of $\lambda_S$ and $g_s$ that is of
interest to us, we can proceed by pretending that it is diagonal.

This description also allows us to interpret the choice of UV cut-off as a choice
of integration measure in the path integral description. 
Let us denote by $\Phi$ the set of all bosonic fields in the theory and let us for
simplicity assume that in an appropriate gauge
the quadratic term
in the fluctuating fields in the action has the form:
\be
S^{(2)}=-{1\over 2} \int d^4 x\,  \Phi.K. \GG^{\mu\nu} D_\mu D_\nu \Phi\, ,
\ee
where $K$ is some matrix determined by the background fields 
that contracts the tensor indices of $\Phi$. It also
includes the $\sqrt{\det \GG}$ factor in the integrand. Then we define the path
integral measure such that
\be\label{emeasurex}
\int [D\Phi] \exp\left[ -{1\over 2} \int d^4 x\,  \Phi.K \Phi\, 
\right] =1\, .
\ee
In this case the path integral
\be 
\int [D\Phi] \exp[-S^{(2)}]
\ee
will produce the desired result $(\det'[-\GG^{\mu\nu} D_\mu D_\nu])^{-1/2}$ after
removing the zero mode contribution. For fermions the prescription is similar,
with $\GG^{\mu\nu} D_\mu D_\nu$ replaced by $\wt\EE_a^\mu D_\mu$.
The advantage of representing the UV cut-off  as a choice of path integral 
measure is that we can now use the same measure
\refb{emeasurex} to determine the integration measure over the zero modes.

We shall illustrate the procedure by computing the integration measure over the metric.
Let us label the fluctuation of string frame metric $G_{\mu\nu}$ as
\be
G_{\mu\nu} = G_{B\mu\nu} +  h_{\mu\nu}\, ,
\ee
where the subscript $B$ stands for background fields.
From now on we shall drop
the subscript $B$ with the understanding that $G_{\mu\nu}$ and $\GG_{\mu\nu}$
will refer to the background metrics $\lambda_S^2 g^{(0)}_{\mu\nu}$ and
$\lambda_S^2 g_s^{-2\gamma} g^{(0)}_{\mu\nu}$ respectively.
Then in the de Donder gauge 
the quadratic action for
$h_{\mu\nu}$ has the form
\be\label{e5.18}
-{1\over 2} \int d^4x \, e^{-2\phi}\, \sqrt{\det G} \, G^{\mu\rho}
G^{\nu\sigma} h_{\mu\nu}\,  {G }^{\alpha\beta}  D_\alpha D_\beta \, h_{\rho\sigma}\, .
\ee
Using \refb{eggdef} and \refb{estring}, this can be written as
\be
 - {1\over 2} \int d^4x \, e^{-2(1+\gamma) \phi}\, 
 \sqrt{\det g^{(0)}} \,g^{(0)\mu\rho}
g^{(0)\nu\sigma} 
 h_{\mu\nu}  {\GG }^{\alpha\beta} D_\alpha D_\beta h_{\rho\sigma}\, .
\ee
Hence the integration measure over
$h_{\mu\nu}$ needs to be normalized as
\be\label{e5.21}
\int [Dh_{\alpha\beta}] \, \exp\left[-{1\over 2}\int d^4x \, \sqrt{\det g^{(0)}}
\, g_s^{-2(1+\gamma)} \,g^{(0)\mu\rho}
g^{(0)\nu\sigma} 
 h_{\mu\nu}  h_{\rho\sigma}\right]=1\, .
 \ee
 Alternatively we can say that $[Dh_{\mu\nu}]$ represents the path integral over the
 variables
 $g_s^{-(1+\gamma)}\,  h_{\mu\nu}$ with no additional $g_s$ and $\lambda_S$
 dependent factors. Similar analysis can be carried out for all other fields.
 
 We shall now describe how this formalism can help us determine the $\lambda_S$ 
 and $g_s$ dependence of the integral over the zero modes.
 First let us focus on the translational zero modes. These can be
regarded as diffeomorphisms with  parameters  that do not
vanish at infinity, but the metric deformation 
generated by these diffeomorphisms vanish at infinity and are allowed deformations. 
Let us call the parameters associated with the three 
diffeomorphisms
as $\alpha^{(i)}$ with $1\le i\le 3$
such that they describe the translation of the black hole 
along the three spatial directions. 
Then each $\alpha^{(i)}$ will have a range $L/\lambda_S$
since $L$ is the
length of the confining box measured in the string metric and since asymptotically 
the string metric approaches $\lambda_S^2 \delta_{\mu\nu}$.
Therefore once we determine the integration measure for the $\alpha^{(i)}$'s,
we shall be able to evaluate the result of integration over the zero modes.

The integration measure over the $\alpha^{(i)}$'s is found as follows.
Deformation by the parameter $\alpha^{(i)}$ corresponds to diffeomorphism by a
vector field $\alpha^{(i)}
\xi^{(i)\mu}(x)$ such that asymptotically $\xi^{(i)\mu}(x)$ approaches
$\delta_{i\mu}$ and deep in the interior $\xi^{(i)\mu}(x)$ vanishes. 
Now under a diffeomorphism of the original background 
with parameter $\xi^\mu(x)$, we generate a metric deformation $h_{\mu\nu}$
of the form
\be
h_{\mu\nu}
=D_\mu (G_{\nu\rho}\, \xi^\rho) + D_\nu (G_{\mu\rho}\, \xi^\rho)\, .
\ee
Using the fact that $G_{\mu\nu}$ scales
as $\lambda_S^2$, we see that
\be
h_{\mu\nu} \sim \lambda_S^2 \alpha^{(i)}\, .
\ee
Now we have already seen that the path integral over $h_{\mu\nu}$ is 
normalized so that it
corresponds to integrating over $g_s^{-(1+\gamma)} h_{\mu\nu}$. 
Therefore measure of integration over the variable $\alpha^{(i)}$ is such that it
corresponds to integration over the variable $g_s^{-(1+\gamma)}\lambda_S^2 
\alpha^{(i)}$. Using the fact that integration over each $\alpha^{(i)}$ spans a range
$L/\lambda_S$, we see that the net contribution from the three translational zero modes
takes the form
\be \label{e5.22y}
\{g_s^{-(1+\gamma)}\lambda_S \, L\}^3\, .
\ee

Analysis of the contribution from the rotational zero modes is similar with the only
difference that now the analog of the parameter $\alpha^{(i)}$ are the rotation angles and
they have range of order unity. This gives the contribution from the two rotational zero
modes to be
\be\label{e5.23y}
\{g_s^{-(1+\gamma)}\lambda_S^2 \}^2\, .
\ee

Next we turn to the contribution from the fermion zero modes. Again, let us assume
for simplicity that in an appropriate gauge the quadratic part of the
action can be written as
\be
S^{(2)} = \lambda_S^\alpha g_s^\beta \int d^4 x \, \Psi . K^{(0)} . \wt\EE_a^\mu D_\mu \Psi\, ,
\ee
where $K^{(0)}$ has no dependence on $\lambda_S$ and $g_s$.
Then $\wh\Psi = \lambda_S^{\alpha/2} g_s^{\beta/2}\Psi$ will have the $\lambda_S$ and
$g_s$ independent integration measure in the sense that the path integral measure
can be expressed as
\be
\int \prod_i d\wh\Psi_i(x) ,
\ee
where $\wh\Psi_i$ are the components of $\wh\Psi$.

Now zero modes are the modes of $\wh\Psi$ for which $S^{(2)}$ vanishes and hence
the integration over these modes would give zero. This is avoided by the insertion of
the $(J_3)^n$ factor in the path integral. It follows from the usual Noether prescription
that $J_3$ has the form
\be
J_3=\lambda_S^\alpha g_s^\beta \int d^3 x \, \Psi . K^{(0)} . \wt\EE_a^0  M \Psi
= \int d^3 x \, \wh\Psi . K^{(0)} . \wt\EE_a^0  M \wh\Psi\, ,
\ee
where $M$ is a matrix of order unity,
measuring the $J_3$ eigenvalues of the components of $\Psi$.
Since $\wt\EE_a^0$ is of order $\lambda_S^{-1} g_s^{\gamma}$, every insertion
of $J_3$ comes with a factor of $\lambda_S^{-1} g_s^{\gamma}$ and two factors of
$\wh\Psi_i$. Therefore we have altogether $2n$ factors of $\wh\Psi_i$ and
a factor of $\lambda_S^{-n} g_s^{n\gamma}$. The integration over the $2n$ zero modes
of $\wh\Psi$ now removes the $2n$ factors of $\wh\Psi_i$ and we are left with
\be \label{e5.27y}
(\lambda_S^{-1} g_s^\gamma)^n = \lambda_S^{-n} g_s^{n\gamma}\, .
\ee
Note that $J_3$ also has contribution from non-zero modes. But if we pick that contribution
from one or more factors of $J_3$ in $(J_3)^n$, we shall miss the corresponding zero mode
contribution and in the expansion of $(J_3)^n$ we shall not have $2n$ factors of
fermion zero modes that are required to saturate the integration over the zero modes.

This takes into account all the zero modes that appeared in the counting \refb{enz}.
Note however that \refb{enz} was obtained under the assumption that the only fields
in the theory are scalars, vectors, metric and spin 1/2 and spin 3/2 fermions. In perturbative
string theory we have an NSNS 2-form field that has been dualized to a scalar in the
analysis leading to \refb{enz}. As already stated in the paragraph below \refb{entot},
the non-zero mode contribution to the path integral does not change under duality
transformation. Since the total number of modes was computed using the description
where we have scalars but no 2-forms, it was appropriate to count the zero modes by
regarding the fields as scalars, so that the number of
non-zero modes is counted correctly. However, since we are now evaluating the zero
mode contribution, we need to treat the field as a 2-form field and count its zero modes.

So we consider the integration over the
NSNS 2-form field $B_{\mu\nu}$. The first part of the analysis proceeds in the
same way as for the metric, and we conclude, as below \refb{e5.21}, that the
integration variable with $g_s$ and $\lambda_S$ independent measure is
$g_s^{-1-\gamma} B_{\mu\nu}$. The zero modes of $B_{\mu\nu}$ are the
two normalizable harmonic 2-forms in the background \refb{eg0}, which can
be taken to be proportional to the electromagnetic field strengths of electrically and
magnetically charged Kerr-Newman solution respectively. 
We shall expand the 2-form field as
\be
B = \sum_{k=1}^2 b_k B^{(k)}, 
\ee
where $B^{(k)}$ are the harmonic forms, normalized so that in the coordinate
system used in \refb{eg0}, they do not carry
any factors of $\lambda_S$ and $g_s$. Then it follows from the result
for the integration measure over $B_{\mu\nu}$ that the variables 
$g_s^{-1-\gamma} b_k$ have $g_s$ and $\lambda_S$ independent integration
measure.

The remaining task is to find the integration range of the variables $b_k$. 
For this we note that 
the fundamental string
world-sheet path integrand has a term proportional to $e^{i\int_\SSS B}
= e^{i \sum_k b_k \int _\SSS B^{(k)}}$ where 
$\SSS$ is the image of the world-sheet in space-time. 
Since $B^{(k)}$'s have been chosen so that in the coordinate system used
in \refb{eg0} they do not carry any factor of $\lambda_S$ and $g_s$, and the
ranges of these coordinates do not have any dependence on 
$\lambda_S$ and $g_s$, $\int_\SSS B^{(k)}$ are also independent of
$\lambda_S$ and $g_s$. Hence the range of $b_k$ are also independent of
$\lambda_S$ and $g_s$. 
Therefore, integration over $g_s^{-1-\gamma} b_k$ for $k=1,2$
is of order
\be \label{e5.29y}
g_s^{-2(1+\gamma)}\, .
\ee

Given that in the NSNS sector one dualizes the axion field into a 2-form field, one may
wonder if the same should be done for the scalars coming from the RR sector. To
this end, note that for the non-zero mode contribution there is no change in the
contribution after a duality transformation. For the zero modes, we recall that the zero
mode deformation are pure gauge with gauge transformation parameters that do not
vanish at infinity. Indeed the zero modes of the 2-form field can be locally written as
$d\Lambda$ with the functional form of $\Lambda$ being the same as the electric 
and the magnetic 4-vector potentials 
associated with the electrically and magnetically charged 
Kerr-Newman
black holes.
Hence the field strengths associated with the zero mode deformations
vanish. On the other hand for the RR fields the path integral is carried out over the 
field strengths instead of gauge fields. This is apparent from string field theory where the
interacting part of the theory involves RR fields in the $(-1/2,-1/2)$ picture, and in this sector
we only see the RR field strengths. Hence the zero modes are not part of the integration
variables in the path integral and we do not need to worry about them.

Taking the logarithm of the product of \refb{e5.22y}, \refb{e5.23y}, \refb{e5.27y} 
and \refb{e5.29y} 
we get the zero mode contribution to $\Delta \ln Z_n$:
\be \label{ezerox}
\Delta_z \ln Z_n = 3\, \ln L + (7-n)\ln\lambda_S -(7+7\gamma - n\gamma) \ln g_s\, .
\ee

\subsection{Total contribution to $S_{BPS}$}

We can now add \refb{endiffx} and \refb{ezerox} and substitute into
\refb{efinfor} to get the total
logarithmic contribution to $S_{BPS}$:
\be\label{efintotx}
S_{BPS} = \cases{ S_{cl}  
-8\ln \lambda_S +(n_{NS}-4)\ln g_s 
+8\gamma\ln g_s \quad \hbox{for $\NN=8$}\cr
S_{cl}  
-4\ln \lambda_S +(n_{NS}-4)\ln g_s 
+4\gamma\ln g_s \quad \hbox{for $\NN=6$}\cr 
S_{cl}  
 +(n_{NS}-4)\ln g_s 
 \quad \hbox{for $\NN=4$}\cr
S_{cl}   +{1\over 6}(23-n_V+n_H) \ln \lambda_S 
+(n_{NS}-4)\ln g_s - {1\over 6}\gamma(23-n_V+n_H) \ln g_s
 \quad \hbox{for $\NN=2$}.}
\ee

\sectiono{Why the UV cut-off should be string scale} \label{s6}

In this section we shall discuss some applications of \refb{efintotx}. The main goal
of this section will be to demonstrate that the gravitational path integral can give us
additional information beyond what is possible from infrared analysis if we use
the fact that the UV cut-off is given by the string scale, without having to do a full
fledged string (field) theory calculation. Of course getting the exact
one loop result will require
us to do a full one loop string theory calculation.

\subsection{Black holes carrying pure NSNS charges} \label{s6.1}

In this section we shall 
consider a BPS black hole carrying pure NSNS charges\cite{2605.30417}. 
Let $Q$
be the electric charge vector and $P$ be the magnetic charge vector, scaling as
$\lambda_Q$ and $\lambda_P$ respectively. Then in Planck units, the mass of the
black hole, given by the BPS formula, takes the form:
\be\label{ezx1}
m(Q,P)^2  \sim c_1 \lambda_Q^2 g_s^2 + c_2 \lambda_P^2 g_s^{-2} 
+ c_3 \lambda_Q \lambda_P\, ,
\ee
where $c_1$, $c_2$ and $c_3$ are $g_s$ independent but could depend on the other
moduli fields.
The attractor value\cite{9508072} 
of $g_s$ is given by extremizing $m$ with respect to $g_s$. This gives
\be\label{ezx2}
g_s \sim \sqrt{\lambda_P/\lambda_Q}\, ,
\ee
where $\sim$ means equality up to a $g_s$ independent multiplicative factor. 
As already discussed, we consider black  hole solutions where the moduli are set
equal to their attractor values so that they do not vary in the solution. So if we
take $\lambda_Q>>\lambda_Q$ then the string coupling $g_s$ is small everywhere
in the solution and we can
trust the perturbation expansion.

In Planck units the overall size of the black hole is of the same order as the mass of the
black hole. From \refb{ezx1} and \refb{ezx2} we see that this is given by
$\sqrt{\lambda_Q\lambda_P}$. Hence the size of the black hole, measured in string
units, is given by,
\be\label{ezx3}
\lambda_S \sim g_s \, m(Q,P) \sim \lambda_P\, .
\ee
We shall consider the case $\lambda_Q>>\lambda_P>> 1$ so that string coupling
is small and the size measured in string units is large. In this case both the
stringy corrections and the quantum corrections are small.

In order to apply \refb{efintotx} we also need to compute $n_{NS}$. These come
from components of the metric and 2-form fields along the internal circles. To this end
we shall consider the following theories:
\begin{enumerate}
\item $\NN=8$ supersymmetric theory: 
This comes from type II string theory on $T^6$ times the (3+1)
dimensional Minkowski space $R^{3,1}$. Since there are 6 internal circles, there are
altogether 12 NS sector gauge fields. Substituting this and \refb{ezx2} and \refb{ezx3}
into \refb{efintotx} we get
\be \label{ezx4}
S_{BPS} = S_{cl} - 4\ln (\lambda_Q\lambda_P) - 4\gamma\ln(\lambda_Q/\lambda_P)\, .
\ee
The exact result for $S_{BPS}$ in this case is known from microscopic counting and
it agrees with \refb{ezx4} for $\gamma=0$. This shows the need for choosing the
UV cut-off to be the string scale.

\item $\NN=6$ supersymmetric theory: This theory is constructed by taking an 
asymmetric orbifold of type IIA theory on $T^4\times T^2\times R^{3,1}$ where the orbifold
group acts on the left-moving world-sheet fields associated with $T^4$ and as a
translation along $T^2$\cite{9508064}. The result is
that among the 8 NS sector gauge fields associated  with $T^4$ in the parent theory,
only 4 survive the orbifold projection. All 4 NS sector fields associated with $T^2$
also survive the projection. Therefore we have $n_{NS}=8$.
Substituting this and \refb{ezx2} and \refb{ezx3}
into \refb{efintotx} we get
\be \label{ezx5}
S_{BPS} = S_{cl} - 2\ln (\lambda_Q\lambda_P) - 2\gamma\ln(\lambda_Q/\lambda_P)\, .
\ee
In this case the microscopic result is not known. Since in string theory we expect $\gamma$
to vanish, \refb{ezx5} gives the following prediction for the microscopic result 
in the limit
$\lambda_Q>>\lambda_P>>1$:
\be \label{ezx5a}
S_{BPS} = S_{cl} - 2\ln (\lambda_Q\lambda_P) \, .
\ee

\item $\NN=4$ supersymmetric theories: There are a wide class of $\NN=4$ supersymmetric
theories. 
There are different equivalent descriptions of these theories related by duality 
transformations, but in this subsection we shall
use the ones that contain only NS sector fields. These include heterotic string theory
on $T^6\times R^{3,1}$ and its asymmetric orbifolds with the orbifold group acting on the
left-moving sector of the theory\cite{9505054,9506048}. 
This also includes asymmetric orbifolds of type II theories
on $T^6\times R^{3,1}$ with the orbifold group acting on the left-moving 
sector, projecting out all the supersymmetry generators arising in the left-moving
sector of the world-sheet\cite{0607155,0609109}. 
$n_{NS}$ differs between different theories, but in each
case $n_{NS}$ is the total number of $U(1)$ gauge fields since all the possible U(1)
gauge fields arising from the RR sector are projected out. 
Substituting \refb{ezx2} and \refb{ezx3}
into \refb{efintotx}, we now get
\be \label{ezx6}
S_{BPS} = S_{cl} -{1\over 2} (n_{NS}-4)\ln (\lambda_Q/\lambda_P) \, .
\ee
The exact results for $S_{BPS}$ in these theories are known from 
microscopic counting\cite{9607026,0605210,0607155,0609109} and in all cases
these results agree with \refb{ezx6} in the 
$\lambda_Q>>\lambda_P>>1$ limit\cite{2605.30417}.

\item $\NN=2$ supersymmetric theory: $\NN=2$ supersymmetric theories with one or
more NSNS sector gauge fields are obtained from heterotic string theory on 
$K3\times T^2$ or some of its orbifolds that do not break any additional 
supersymmetries. 
In this case we are left with four NS sector gauge fields associated with the 
$T^2$
directions and some more from the ten dimensional gauge fields whose precise
number depends on the choice of the gauge bundle on $K3$. 
However since all the gauge fields come from the NS sector, we have
$n_{NS}=n_V+1$, where $+1$ comes from the graviphoton field.
Substituting $n_{NS}=n_V+1$, and 
\refb{ezx2} and \refb{ezx3}
into \refb{efintotx} we get
\be \label{ezx7}
S_{BPS} = S_{cl} +{1\over 6}(23-n_V+n_H) \ln \lambda_P - {1\over 2}
(n_V-3)\, \ln (\lambda_Q/\lambda_P)
 + {1\over 12}\gamma(23-n_V+n_H) \ln (\lambda_Q/\lambda_P)\, .
\ee
In this case the microscopic result is not known. Since in string theory we expect $\gamma$
to vanish, \refb{ezx7} gives a prediction for the microscopic result in the limit
$\lambda_Q>>\lambda_P>>1$:
\be \label{ezx8}
S_{BPS} = S_{cl} +{1\over 6}(23-n_V+n_H) \ln \lambda_P - {1\over 2}
(n_V-3)\, \ln (\lambda_Q/\lambda_P)
 \, .
\ee
\end{enumerate}

\subsection{Black holes carrying pure RR charges} \label{s6.2}

We now consider black holes carrying RR charges with all the RR charges scaling as
$\lambda$\cite{2606.17149}. In Planck units, the mass of such a black hole is
\be
m = c\lambda\, ,
\ee
where $c$ is a function of various moduli other than the dilaton.
Since there is no dependence on the string coupling $g_s$, it is not fixed by the
attractor equation and can take an arbitrary value. In Planck units the black hole size
is of the order of its mass of order $\lambda$. Hence the black hole size in string
units is given by
\be
\lambda_S \sim \lambda \, g_s\, .
\ee
Substituting this into \refb{efintotx} we get
\be\label{efintotxkk}
S_{BPS} = \cases{ S_{cl}  
-8\ln \lambda +(n_{NS}-12)\ln g_s 
+8\gamma\ln g_s \quad \hbox{for $\NN=8$}\cr
S_{cl}  
-4\ln \lambda +(n_{NS}-8)\ln g_s 
+4\gamma\ln g_s \quad \hbox{for $\NN=6$}\cr 
S_{cl}  
 +(n_{NS}-4)\ln g_s 
 \quad \hbox{for $\NN=4$}\cr
S_{cl}   +{1\over 6}(23-n_V+n_H) \ln \lambda
+\left\{n_{NS} + {1\over 6}(-1-n_V+n_H)\right\} \ln g_s \cr
\hskip 2in - {1\over 6}\gamma(23-n_V+n_H) \ln g_s
 \quad \hbox{for $\NN=2$}.}
\ee

Since $g_s$ is arbitrary, $S_{BPS}$ should be independent of $g_s$.
Using $n_{NS}=12$ for $\NN=8$ and $n_{NS}=8$ for $\NN=6$, we see that
$S_{BPS}$ is indeed independent of $g_s$ if we set $\gamma=0$. For $\NN=4$
the situation is a bit more subtle since in the description of the theories given in
section \ref{s6.1} all the RR sector U(1) fields are projected out and we cannot get
a black hole carrying RR charges. To remedy this we use a dual description of these
theories as (orbifolds of) type II theory on $K3\times T^2\times R^{3,1}$ where the orbifold group
acts as a translation along $T^2$ and a holonomy preserving transformation on $K3$
or an orbifold of type II on $T^4\times T^2\times R^{3,1}$ where the orbifold group
acts as a translation along $T^2$ and a rotation along $T^4$
producing $SU(2)$ holonomy. As a result in both classes of theories the only NS sector
U(1) gauge fields are those coming from the components of the metric and the Kalb-Ramond
field along $T^2$ and we have $n_{NS}=4$. Therefore we again see that the $g_s$
dependence disappears.

$\NN=2$ supersymmetric theories containing RR U(1) gauge fields come from type IIA or
IIB string theory on $CY_3\times R^{3,1}$. In this case we have $n_{NS}=0$, since there
are no internal circles. We now see that even for $\gamma=0$ the coefficient of the
$\ln g_s$ term does not vanish. Instead we have a contribution
\be \label{e6.13x}
{1\over 6}(-1-n_V+n_H)\, \ln g_s =\pm {\chi_{CY}\over 12} \ln g_s\, ,
\ee
where $\chi_{CY}$ is the Euler number of $CY_3$ and the + and $-$ signs are for IIB 
and IIA theories respectively.
This produces an apparent contradiction since we expect the index to be $g_s$
independent. 

This contradiction could be avoided if the one loop effective action contained a term
$\mp {\chi_{CY}\over 24} \ln g_s\, E_4$ where $E_4$ is the Euler density in four
dimensions. In that case, since the Euclidean black hole has Euler number 2,
the effective action would evaluate to $\mp {\chi_{CY}\over 12} \ln g_s$ which would
cancel the unwanted term in \refb{e6.13x}.\footnote{Our sign convention for the
Euclidean action $\SSS$ is such that the path integral is weighted by $e^\SSS$.} 
The problem with this is that since $g_s$ is
the loop expansion parameter, we expect the genus $g$ contribution to be proportional
to $g_s^{2g-2}$. According to this logic the one loop amplitude should  be $g_s$
independent. It turns out that this argument breaks down due to the existence of an infinitely
degenerate tower of states in the RR sector of type II string theory. The existence of this
tower makes the naive one loop result diverge. We can avoid this issue by computing the
derivative of the partition function with respect to $g_s$ and then integrating this with
respect to $g_s$. This procedure produces a term $\mp {1\over 24} \ln g_s\, E_{10}$
where $E_{10}$ is the ten dimensional Euler density\cite{2607.12031}. 
Upon compactification on $CY_3$
times the black hole space-time
this produces the desired term that cancels the $g_s$ dependent term in \refb{e6.13x}.
This term does not affect the results for the $\NN\ge 4$ theories, since the Euler number
of the internal space vanishes in these theories. 

\sectiono{Does supergravity need stringy cut-off?} \label{s7}

In section \ref{s6}  we have discussed why the choice $\gamma=0$, i.e.\
choosing the UV cut-off to be the string scale, seems to be necessary to produce consistent
results for the index. However, some other stringy input also goes 
into the analysis. For example the analysis of section \ref{s6.1} tests agreement with
the microscopic results that come from string theory. The analysis of section \ref{s6.2} does
not rely on any microscopic result but uses the fact that the index should be independent
of the moduli that are not fixed by the attractor equation. This is needed for internal
consistency independent of UV completion. Nevertheless, some stringy input was used,
{\it e.g.} the use of the NSNS 2-form field instead of the axion as path integral variable,
dropping integration over the zero modes of the RR field etc. In this section we shall
examine whether we can reach the conclusion that the UV cut-off is the string scale
if we do not use these stringy inputs and / or whether these stringy
inputs can be justified by other means.

Now, if we allow the UV completion to generate any term, then we can use them to cancel
the unwanted terms. So 
in order to carry out the analysis we need to fix some criteria that we impose on a possible
UV completion. We use the following criteria:
\begin{enumerate}
\item Given a low energy $\NN\ge 2$ supergravity in (3+1) dimensions, we first carry out
field redefinitions and duality transformation on the fields to express the action as
\be \label{e6x11}
\int d^4 x \, \sqrt{\det G} \, e^{-2\phi} \, \LL\, ,
\ee
where $\phi$ is one of the moduli fields and $\LL$ is invariant under a constant shift
of $\phi$, keeping the other fields fixed. 
In that case the vacuum expectation value of $e^\phi$ can be interpreted as
the coupling constant $g_s$ and we require that for small $g_s$, we shall have a perturbation
expansion in powers of $g_s$ with the $g$ loop contribution being of order $g_s^{2g-2}$.
Any deviation from this  rule must have an explanation based on infrared physics.

Note that in order to express the action in the form \refb{e6x11} we may need to absorb
some factors of $e^\phi$ into the gauge fields so that the gauge transformations
take non-standard form, different from \refb{egauge}. 
This would happen for example for RR fields in string theory
where the natural normalization of the gauge fields for which the gauge transformation
takes the form \refb{egauge} does not produce the $e^{-2\phi}$ term in front of the
gauge field kinetic term. So we need to scale the gauge 
fields by $e^{-\phi}$ to bring the action
to the form \refb{e6x11} and the scaled gauge fields will have different gauge transformation
laws. Also note that the choice of $\phi$ is not unique. Using different
field redefinitions and duality transformations we may be able to bring the action into the
form \refb{e6x11} in more than one way. These correspond to different weak coupling
limits, and the criteria that we impose should hold in
all these limits.

\item There is a specific metric $G_{\mu\nu}$ in terms of which the action takes the form
given in \refb{e6x11}. We shall call it the `string metric' even though we do not assume the
existence of an underlying string theory. The Einstein frame metric $G^E_{\mu\nu}$
is related to the string
metric $G_{\mu\nu}$ by a rescaling,
\be
G_{\mu\nu} = e^{2\phi} \, G^E_{\mu\nu}\, ,
\ee
so that the action, written in terms of the Einstein metric, has no power of $e^\phi$
multiplying the Einstein-Hilbert term in the action.

\item We shall assume that there is a uniform UV length
cut-off, which we take to be $g_s^{\gamma-1}$ times the Planck length or equivalently
$g_s^{\gamma}$ times the would be string length had we embedded the theory into
a string theory. Here Planck length and string length are
defined as the unit length measured in the Einstein metric $G^E_{\mu\nu}$ and
the string metric $G_{\mu\nu}$ respectively. 

\item We shall assume that if there are massive particles present in the theory then
their masses are of the order of the cut-off scale, so that no large logarithms are generated
by the loops of massive particles. This is in the same spirit as the assumption of a
uniform cut-off.

\item We shall assume that the theory contains states charged under the gauge
fields and in particular has charged black holes. The charge lattice can be determined
following the procedure outlined in section \ref{s2}. The choice of the lattice is not
fixed by low energy supergravity but it is fixed in a given theory. The same should hold
for 2-form fields $B_{(\alpha)}$, -- their 
sources are strings carrying the 2-form charge that couple to the
2-form fields via a generalization of \refb{eworld}
\be \label{estringworld}
\sum_\alpha {\mathcal Q}^{(\alpha)} \int_{ws} B_{(\alpha)}\, ,
\ee
where $B_{(\alpha)}$ are normalized so that the gauge transformation acts
as $\delta B_{\alpha)}= d\Lambda^{(\alpha)}+\cdots$ with the $\cdots$ not having
any $\Lambda^{(\alpha)}$ without derivatives. We shall call this the 
'natural normalization'.
Then it follows that ${\mathcal Q}^{(\alpha)}$'s take
value on a fixed lattice independent of the moduli fields. Note that this
natural  normalization
of the $B_{(\alpha)}$'s could differ from the one that brings the action to the form
\refb{e6x11}, -- this is what happens for the RR fields. Similar convention for
natural normalization can be adopted for other form fields, including gauge fields.

Note that the existence of a string that couples to a 2-form field is not entirely
an ad hoc criterion. In the absence of such a string the zero mode of the 2-form
field will not have well defined period. If we regard the absence of a string as the limit
when the minimal charge carried by the string becomes infinite, then in this limit
the period of the zero mode will vanish and 
the integration over such a zero mode will
make the partition function $Z_n$ vanish. Since all Euclidean black holes have
2-form zero modes, none of them would contribute to the path integral in such a
theory.

\item Once we have brought the action to the form given in \refb{e6x11} we may have
a choice of making a duality transformation followed by rescaling of fields to preserve
the form of \refb{e6x11}. In that case we need to decide which are the fundamental field
variables over which the path integral needs to be performed. This is done by demanding
that when we use natural normalization for the form fields, 
then the kinetic term of a fundamental field, 
written in the metric $G_{\mu\nu}$ that appears in
\refb{e6x11}, will have a multiplicative factor of $e^{-\alpha\phi}$ with $\alpha\ge 0$ so that
it is weakly coupled.
Then the kinetic term for the dual field will have a multiplicative factor of $e^{\alpha\phi}$.
This would for example single out the
NSNS 2-form field as the fundamental field since 
its kinetic term is accompanied by a factor of $e^{-2\phi}$. The dual axion, on the other hand,
has a kinetic term with a multiplicative factor of $e^{2\phi}$ and is strongly coupled.
This will, however, not tell us whether
we should use the RR axion or the 2-form field since we have $\alpha=0$ for both, i.e.\
in the natural normalization both kinetic terms are free from any multiplicative factor
involving $\phi$. We shall address this problem shortly.

\item If we can construct a black hole for which $g_s$ is not fixed by the 
attractor mechanism,
then the index of the black hole, computed by the procedure described in section
\ref{s4}, should be independent of $g_s$ for a
fixed charge vector on the lattice. This in fact should be a general property,
-- if the attractor mechanism does not fix a modulus, then the index should be
independent of that modulus, -- but we shall focus on $g_s$. The logic for this
is that we could produce a background where the moduli vary slowly and then
let the black hole roll in this background, and the index should remain unchanged
during the roll. However we {\em should not} 
demand that the index remain unchanged when
we change the parameters of the charge lattice keeping the integers labelling the charges
fixed.

\end{enumerate} 

Given these criteria, the only ambiguity left over from the analysis of section \ref{s6.2}
is in the choice of using 0-form field or the dual 2-form field as the fundamental field
in the RR sector, -- where by RR sector we shall refer to sectors
where the kinetic terms, in the natural normalization of the field variables, do not have
the $e^{-2\phi}$ factor in the action. The analysis in section \ref{s6.2} effectively used the
0-form fields as the fundamental fields, since we had said that even if we use 2-form fields,
their zero modes  are not integrated over in string theory. But if we do not use the insight
from string theory, and use the 2-form field as the fundamental field, we would integrate
over their zero modes. The contribution from these integrals can be evaluated in the same
way as in the case of NSNS 2-form fields, except that due to the absence of the $e^{-2\phi}$
factor in the kinetic term,  the $g_s^{-2(1+\gamma)}$ factor in \refb{e5.29y} will now
be replaced by $g_s^{-2\gamma}$. Hence if we convert $m_R$ of the scalars to 2-forms,
we shall get a factor of $g_s^{-2 m_R\gamma}$ in $Z_n$, leading to an
additive contribution of $-2m_R\gamma\ln g_s$ in $\ln Z_n$. 
Augmenting this by \refb{efintotx} we see that  the net $\gamma$
dependent terms in $S_{BPS}$ will be of the form:
\be\label{egammadep}
\Delta_\gamma S_{BPS} = \cases{ (8-2m_R)
\gamma\ln g_s \quad \hbox{for $\NN=8$}\cr
(4-2m_R)\gamma\ln g_s \quad \hbox{for $\NN=6$}\cr 
-2m_R \gamma\ln g_s 
 \quad \hbox{for $\NN=4$}\cr
 - {1\over 6}\gamma(23-n_V+n_H) \ln g_s - 2m_R\gamma\ln g_s
 \quad \hbox{for $\NN=2$}.}
\ee
Since the $\gamma$ independent terms have already been shown to give a result
for $S_{BPS}$ that is independent of $g_s$, we require, for consistency, that 
the right hand sides of \refb{egammadep} vanish. 
Now, in $\NN=8$ supergravity we have 32 RR sector scalars, and there is no
intrinsic difference between them.\footnote{For example under the continuous duality
group $SO(6,6)$ that preserves the weak coupling limit $e^\phi<<1$, the 32 scalars
transform in the spinor representation.} Thus the only natural possibilities are
to either dualize all the scalars or not dualize any of them. This gives $m_R=32$ or 0,
and in either case $\gamma$ needs to vanish in order for the right hand side of
\refb{egammadep} to vanish. For $\NN=6$ theory we have 16 RR sector scalars and
the natural choice of $m_R$ is 0 or 16. So again $\gamma$ needs to vanish for the
right hand side of \refb{egammadep} to vanish.
Hence for $\NN=6,8$ supergravity, we need the UV cut-off to be of the order of 
the string scale. 
For $\NN=4$ supergravity vanishing of the right hand side of  \refb{egammadep}
forces us to have $m_R=0$
or $\gamma=0$. Hence if we
decide to work with RR 0-form fields as fundamental fields (or use 2-form fields as
fundamental fields but ignore integration over its zero modes) then the result is
insensitive to the choice of UV cut-off.

The case of $\NN=2$ supergravity is somewhat involved since even for $\gamma=0$
we have extra term proportional to $\ln g_s$ as given in \refb{e6.13x} that needs to be
added to the effective action. According to our criteria for UV completion, all such extra
terms must have infrared origin. So we can examine if \refb{e6.13x} has infrared origin.
To this end we note that these extra terms arise because in the underlying string field theory
we have infinite degeneracy at each mass level. Such infinite degeneracy is present even
at the massless level; if we take the $-1/2$ picture R sector massless states 
and apply the zero
mode $\gamma_0$ of the $\gamma$ ghost field, we generate a state with the same 
$L_0$ eigenvalue. This can be repeated arbitrary number of times generating 
arbitrary number of $L_0=0$ states. In physical amplitudes this effect is tamed by the
R-sector propagator that only allows a finite number of these states to 
propagate\cite{1703.06410}, but the
partition function diverges. Since the effect is visible at the massless level, it is an infrared
effect, although of an unusual type since it is not visible at the level of supergravity.

Since for $\gamma=0$ the $g_s$ dependent terms cancel after taking into account the
effect of \refb{e6.13x}, the right hand side of \refb{egammadep} must vanish for $\NN=2$
as well. Depending on the value of $n_V$ and $n_H$, this may be possible for some choice
of $m_R$ as long as that choice is a non-negative integer not exceeding the number
of RR scalars. However there is no universal
solution independent of $n_V$ and $n_H$ other than $\gamma=0$ even in this case.

\sectiono{The way forward} \label{s8}

From the analysis described above, it seems clear that if we want to do a precise computation
of any quantity using gravitational path integral, we need the knowledge of UV completion
of the theory. At the level of computation described in this review, we did not need the
full string theory, -- only the knowledge of stringy cut-off and the knowledge of
correct duality frame for
the elementary fields was sufficient. However, if we want to go beyond this order, we shall
need the full world-sheet formalism to compute the one loop (and higher loop) contribution
to the partition function. 
At the same time, we cannot forget about the gravitational path integral description altogether,
since that is needed to properly take into account the effect of the IR cut-off and separate
the contribution due to the modes attached to the black hole from the continuum.
String field theory is a natural framework that combines the two approaches, and it could
be useful for a systematic computation of the macroscopic entropy / index.

One could nevertheless worry that since we do not have an exact conformal
field theory description of the black hole solution and / or the black hole carries
flux of RR fields, we may not be able to formulate string (field) theory in the black hole
background. Note however that in the large charge limit the fields are weak and hence we
could expect that the techniques developed in 
\cite{1811.00032,2311.04959,2507.12921} may be useful.

One could also explore the possibility of combining localization techniques with the
world-sheet approach. This approach will be less ambitious, in the sense that we do
not attempt to get all loop contribution at one go. Instead we use localization to 
simplify the analysis at each order in the string loop expansion.

Finally, one would like to extend the analysis to compute quantum corrections to 
non-supersymmetric black hole entropy. The techniques used here are the same techniques
that were used in the analysis of logarithmic corrections to non-supersymmetric black hole
entropy\cite{1205.0971}, and hence we might 
expect that with some extra effort we should be able to
take into account the effect of stringy corrections to the entropy of non-supersymmetric
black holes.

\bigskip

\noindent{\bf Acknowledgement:} 
I wish to thank A.~H.~Anupam, P.~V.~Athira, Chandramouli Chowdhury, 
Muktajyoti Saha and P. Shanmugapriya for discussions
and collaboration in various papers
whose results
were used in this paper. I also thank 
Sergey Alexandrov, Subramanya Hegde, 
Luca Iliesiu, Don Marolf, Sameer Murthy, 
Boris Pioline, Mukund Rangamani, Victor Rodriguez, 
Joaquin Turiaci and Amitabh Virmani 
for useful communications on various topics discussed in this review.
I thank P.~Shanmugapriya and Amitabh Virmani for discussions and
for critical comments on an earlier version of appendix \ref{sa} of the manuscript.
This work was supported by the ICTS-Infosys Madhava 
Chair Professorship and
Department of Atomic Energy, Government of India, under project no.~RTI4019.

\appendix

\sectiono{New Attractor Solution at the Attractor Point} \label{sa}

In this appendix we shall analyze the new attractor solution when the asymptotic
values of the scalars are set equal to their attractor values and show that the scalars
become position independent. We consider a general $\NN=2$ supergravity coupled
to vector multiplets. Hypermultiplets do not couple to vector multiplets at the two
derivative level and remain constant in any black hole solution. The solutions in
$\NN\ge 4$ supergravity can be embedded, after duality rotation, inside a special
$\NN=2$ supergravity known as the STU model, with prepotential $F$ 
proportional to $X^1 X^2 X^3/X^0$.

A solution similar to the one described below has been discussed in 
\cite{2310.07763} but our solution differs from the one in \cite{2310.07763} 
in some minor details,
as will be pointed out in footnote \ref{f8}.

\subsection{New attractor solution} \label{sa2}

The relevant fields that will enter our analysis include the metric $g_{\mu\nu}$ and
a set of complex scalars $\{X^I\}$. 
There are many other fields, including the gauge fields, but for now,
we shall not write down the
solutions for these fields explicitly.
The form of the two derivative 
Lagrangian is controlled by a function $F(\{X^I\})$, known as
the prepotential, which
satisfies the homogeneity relation,
\be\label{ehomx}
F(\{\lambda X^I \}) = \lambda^2 \, F(\{X^I\})\, .
\ee
We shall define:
\be
F_I ={\p F\over \p X^I}\, .
\ee
For now we shall restrict our analysis to two derivative action. We shall comment on the
effect of higher derivative terms later.

Following \cite{0009234}, ref.~\cite{2310.07763} constructed a class of stationary
solutions in this theory preserving
four supercharges. We shall use the notation of 
\cite{2411.08260} to write down the solution, -- more details can be found
in \cite{2511.18771}.
The metric has the form,
\be\label{emetric}
ds^2=-e^{2g}(dt+  \sigma_\alpha \, dx^\alpha)^2+e^{-2g} ds^2_{\mathrm{base}},
\ee
where
\be\label{embase}
ds^2_{\mathrm{base}}={{r^2-b^2\cos^2\theta} \over {r^2-b^2}}dr^2+(r^2-b^2\cos^2\theta)d\theta^2+(r^2-b^2)\sin^2\theta d\phi^2,
\ee
and
\be\label{esigal}
\sigma_\alpha \, dx^\alpha=\omega_\phi \, d\phi\, .
\ee
$g$ and $\omega_\phi$ are smooth functions of $r$ and $\theta$.
We shall collectively denote the base space coordinates
$(r,\theta,\phi)$ by the coordinates $x^\alpha$,
and denote by
$\alpha,\beta,\gamma,\cdots$ the three dimensional curved indices
associated with these coordinates. 
Finally, we define 
\be\label{e226x}
R(\sigma)^\gamma = \ve^{\alpha\beta\gamma} \p_\alpha \sigma_\beta\, .
\ee
Here $\ve^{\alpha\beta\gamma}$ is the epsilon tensor on the three dimensional base
space and the index of $R(\sigma)$ is raised by the three dimensional base
metric \refb{embase}.
The parameter $b$ controls the temperature and angular 
momentum associated
with the black hole solution.  The horizon is situated at $r=b$.
The index $S_{wald}+2\pi i J$ is known to be independent of $b$. 
Since we are studying two derivative action, $S_{wald}$ coincides with the 
Bekenstein-Hawking entropy.

We shall now specify the equations that determine $X^I$, $\sigma_\alpha$ and $g$. 
We shall consider complex solutions by taking $X^I$ and its complex
conjugate $\bar X^I$ as independent variables. $\bar F(\bar X)$ will be obtained
by replacing in the expression for $F(X)$ all the $X^I$'s by $\bar X^I$ and all the $i$'s
by $-i$. $\bar F_I(\bar X)$ will be defined as $\p\bar F(\bar X)/\p\bar X^I$.
We also define,
\be\label{ea2y}
r_N=r-b\cos\theta, \qquad r_S=r+b\cos\theta\, .
\ee
For the solution that is
of interest to us, the fields satisfy the equations:
\ben \label{e242}
-i\, e^{-g} (\bar h X^I - h\bar X^I) &=& \hh^I + {\gamma_N^I\over r_N} + 
{\gamma_S^I\over r_S} \, ,
\nonumber \\
-i\, e^{-g} (\bar h F_I - h\bar F_I) &=&\hh_I + {\gamma_{NI}\over r_N} + 
{\gamma_{SI}\over r_S} \, ,
\een
\ben\label{e232x}
i\, (\bar X^I F_I- X^I\, \bar F_I) +{1\over 2}\chi &=& 0, \qquad \chi=-2 \, ,
\een
\be \label{e233x}
(\bar h \, X^I - h \, \bar X^I) \nablax_\alpha (\bar h \, F_I - h \, \bar F_I)
- (\bar h \, F_I - h \, \bar F_I) \nablax_\alpha (\bar h \, X^I - h \, \bar X^I)
-{1\over 2} \, \chi\, e^{2g}\, R(\sigma)_\alpha = 0\, ,
\ee
\ben \label{e242new}
&&- i\, e^{-g} \bar h X^I  = {\gamma_N^I\over r_N} + \OO(1),\quad 
 - i\, e^{-g}\, \bar h F_I =  {\gamma_{NI}\over r_N}  +\OO(1), 
\qquad \hbox{as $r_N\to 0$}\, ,
\nonumber \\
&& i\, e^{-g}   h\bar X^I = 
\OO(1)\, , \quad 
 i\, e^{-g}  \, h\bar F_I =
\OO(1)\, , 
\qquad \hbox{as $r_N\to 0$}\, ,\nonumber \\
&& i\, e^{-g}   h\bar X^I = 
{\gamma_S^I\over r_S} +\OO(1)\, , \quad 
 i\, e^{-g}  \, h\bar F_I =
{\gamma_{SI}\over r_S} +\OO(1)\, , 
\qquad \hbox{as $r_S\to 0$}\, ,\nonumber \\
&& - i\, e^{-g} \bar h X^I  = \OO(1),\quad 
 - i\, e^{-g}\, \bar h F_I = \OO(1), 
\qquad \hbox{as $r_S\to 0$}\, ,
\een
\be\label{egammasum}
\gamma_N^I +  \gamma_S^I = P^I, \qquad
\gamma_{NI} +  \gamma_{SI} = Q_I\, .
\ee
Here the choice $\chi=-2$ is a gauge choice.
$\hh^I$ and $\hh_I$ are constants determined from the asymptotic values of
$X^I$, $\bar X^I$.
$Q_I$ and $P^I$ are the electric and the magnetic charge vectors carried by the
black hole. $h$ is a position dependent phase and $\bar h$ is its complex conjugate.
%{\bf We can absorb $\bar h$ into $X^I$ and $h$ into $\bar X^I$.}

The way these equations can be solved is as follows. 
Since $F_I$ are known homogeneous functions of degree one in the $X^I$'s, 
the first and the third lines of 
\refb{e242new} give, respectively,
\be
\gamma_{NI} = F_I(\{\gamma_N^J\}), \qquad \gamma_{SI} = \bar F_I(\{\gamma_S^J\})\, .
\ee
This
determines
$\gamma_{NI}$ in terms of $\{\gamma_N^I\}$ and $\gamma_{SI}$ in terms of 
$\{\gamma_S^I\}$. 
$\gamma_N^I$, $\gamma_{NI}$, $\gamma_S^I$ and $\gamma_{SI}$ are then
determined individually
from the equations \refb{egammasum}.
\refb{e242} now determines $e^{-g}\bar h X^I$ and $e^{-g} h \bar X^I$.
$e^{-g}$ is then determined from the equation \refb{e232x}
and $\sigma_\alpha$ is determined from the equation \refb{e233x}.

In order that the metric computed from \refb{e232x}, \refb{e233x}
does not have a Dirac string singularity, we also need to impose
the constraint
\be\label{ehhgam}
 \langle \hh, \gamma_N+\gamma_S\rangle =0\, ,
\ee
where, for any two
vectors $a$, $b$,
\be\label{econx}
\langle a, b\rangle \equiv a^I b_I - a_I b^I\, .
\ee
Given the asymptotic values of the physical scalars, $\hat h^I$ and
$\hat h_I$ are not uniquely fixed
since we can multiply the $X^I$'s by an overall phase. We can use this freedom
to satisfy \refb{ehhgam}.

\subsection{Solution at the attractor point} \label{sa3}

In this section we shall consider a special class of solutions where the asymptotic scalars
are set equal to their attractor values. This corresponds to the choice
\be \label{e3.1x}
\hat h^I = P^I, \qquad \hat h_I =Q_I\, .
\ee
We shall also choose the boundary condition on the metric to be
\be\label{e3.2x}
e^{-g}|_\infty = a\, ,
\ee
for appropriate constant $a$.\footnote{I would like to thank Amitabh Virmani for
pointing out the need to introduce the constant $a$, so that the results are
compatible with the $\chi=-2$ gauge condition for generic choice of charges.}
Normally we take the asymptotic value of $e^{-g}$ to be 1, but for scaling arguments 
it is useful to express the metric in 
such a way that the overall size of the black hole comes out
as an overall multiplicative factor. We shall see that the choice \refb{e3.1x}, \refb{e3.2x}
achieves this after a rescaling of the time coordinate by $a^2$.
Using \refb{egammasum} we see that \refb{ehhgam} holds automatically at the attractor
point for the choice \refb{e3.1x}.
We shall now write down an explicit solution for this choice of the asymptotic values
of the scalars.

First we introduce a set of complex constants $c^I$ defined by
\be\label{etleft}
c^I = -i e^{-g} \bar h X^I|_\infty\, .
\ee
Since the asymptotic values of the moduli satisfy reality conditions, we have
\be\label{etright}
 c^{I*} = i e^{-g} h \bar X^I|_\infty\, .
\ee
It follows from the homogeneity property of the prepotential that 
\be\label{etrighta}
F_I(c) = -i e^{-g} \bar h F_I|_\infty, \qquad F_I(c)^* = i e^{-g} h \bar F_I|_\infty\, .
\ee
Also from the asymptotic form of the equations \refb{e242}, \refb{e232x}, we see
that $c^I$'s satisfy
\be \label{e33a}
c^I+c^{I*}= \hat h^I = P^I \qquad
F_I(c) + \bar F_I(c^*)=\hat h_I =  Q_I\, ,
\ee
and
\be \label{e3.6y}
i\, a^{-2}\,  (c^{I*} F_I(c)- c^I\, \bar F_I(c^*)) =1 \, .
\ee
In arriving at the last equation we have used the boundary condition \refb{e3.2x}. 
\refb{e33a} and \refb{e3.6y} determine $a$ in terms of the charges.

Eqs.~\refb{e3.1x}-\refb{e3.6y} are the same ones that one would use
for the construction of the usual attractor solution describing non-rotating extremal
BPS black holes. We now claim that
the solution to \refb{e242}-\refb{egammasum} with the same boundary 
conditions is as follows:\footnote{Ref.~\cite{2310.07763}
mentioned a similar solution
where $X^I / \sqrt{i(\bar X^J F_J - X^J \bar F_J)}$ remains constant in the solution.
In our solution this ratio is proportional to
$\sqrt{\left(1 +{1\over r_N}\right)\Big/\left(1 +{1\over r_S}\right)}$ 
and is not a constant. The relation
between the solution constructed here and those discussed in 
\cite{2310.07763} is not clear, -- they might be related by a complex
gauge transformation. \label{f8}}
\be  \label{e3.7new}
\gamma_N^I =  c^I, \qquad \gamma_S^I =  c^{I*}, \qquad \gamma_{NI} =
F_I(c), 
\qquad \gamma_{SI} =  \bar F_I(c^*), 
\ee
\be \label{e3.8}
\bar h X^I = \varphi(r,\theta) c^I, \quad h \bar X^I = \bar \varphi(r,\theta) c^{I*}, \quad
\bar h  F_I (X)= \varphi(r,\theta)  F_I(c), \quad
h \bar F_I (\bar X)= \bar \varphi(r,\theta)\bar F_I(c^*)\, ,
\ee
\be \label{e3.9}
e^{-g}\varphi = i\, \left(1 +{1\over r_N}\right), \qquad
e^{-g}\bar\varphi = -i\, \left(1 +{1\over r_S}\right)\, ,
\ee
\be \label{e3.10}
e^{-2g} =a^{2} \left(1 + {1\over r_N}\right)\left(1+{1\over r_S}\right)\, ,
\ee
\be \label{e3.11}
R(\sigma)_\alpha = i\, a^2\, e^{-2g}\, \left\{ \varphi\nablax_\alpha \bar\varphi -
\bar \varphi\nablax_\alpha \varphi \right\}\, .
\ee
Integrating \refb{e3.11} using \refb{e226x}, \refb{esigal} we get
\be \label{eomphixy}
\omega_\phi = -i\, a^2\, b\, \sin^2\theta \, {2\, r +1\over r^2 - b^2\cos^2\theta}\, .
\ee
We can easily verify that this configuration satisfies 
eqs.~\refb{e242}-\refb{egammasum}. Substituting these into \refb{emetric} and
making the replacement $t\to -i\, \tau$,
$\omega_\phi =-i \,  \wt\omega_\phi$ one arrives at the Euclidean solution
given in \refb{eg0}.

For this solution the physical scalars are given by
\be
\psi^I = X^I/X^0 = c^I/c^0, \qquad \bar\psi^I = \bar X^I/\bar X^0 = c^{I*}/c^{0*}\, .
\ee
Since $\bar\psi^I = \psi^{I*}$, the physical 
scalars are constant and satisfy the reality condition  in the solution.

We also see from \refb{e3.10} and \refb{e3.11} that the metric is universal, independent of
the choice of the prepotential $F(X)$ 
and even the number of vector multiplets. Hence the same result
holds when there are no vector multiplets. This corresponds to the case when $I$ takes only
one value 0 and the prepotential is proportional to $i(X^0)^2$. Since there are no physical scalars,
the physical bosonic fields are the metric and a single $U(1)$ field and the solution must
coincide with that of the Euclidean Kerr-Newman metric. Then by the universality of the metric,
the same result continues to hold for general $N=2$ supergravity as long as the asymptotic
values of the scalars are chosen to be the attractor values. This agrees with the 
result of \cite{2605.30417}
for the STU model.

Finally, let us examine the gauge field configurations. First we define
$G_{I\mu\nu}$ via the relation,
\be
\wt G_{I\mu\nu} = {\delta S\over \delta F^{I\mu\nu}}\, ,
\ee
where $\sim$ denotes Hodge dual. Then the result of \cite{0009234} is that
\ben
F^I_{0\alpha} &=& -\nablax_\alpha\left[e^{g}(\bar h X^I + h \bar X^I)\right]\, , \nonumber \\
G_{I0\alpha} &=& -\nablax_\alpha\left[e^{g}(\bar h F_I + h \bar F_I)\right]\, .
\een
Using \refb{e3.8}-\refb{e3.10}, this gives,
\ben
F^I_{0\alpha} &=&- i a^{-2}\, \nablax_\alpha \left[c^I \left(1+{1\over r_S}\right)^{-1} - c^{I*} 
\left(1+{1\over r_N}\right)^{-1}
\right]\, , \nonumber \\
G_{I0\alpha} &=&- i a^{-2}\, 
\nablax_\alpha \left[F_I(c) \left(1+{1\over r_S}\right)^{-1} - F_I(c)^* \left(1+{1\over r_N}\right)^{-1} 
\right]\, .
\een
Hence the gauge fields are also semi-universal, in the sense that both $F^I_{0\alpha}$ and
$G_{I0\alpha}$ are given by linear combinations of $\nablax_\alpha(1+r_N)^{-1}$ and
$\nablax_\alpha(1+r_S)^{-1}$.

Finally, let us comment on the higher derivative corrections to the solution. As was 
shown in
\cite{0009234}, and reviewed in \cite{2411.08260}, higher derivative corrections do not
modify the general form of the metric given in \refb{emetric}, \refb{embase}, \refb{esigal} 
or
eqs.\refb{e242}, \refb{e242new}, but give additional terms in 
\refb{e232x}, \refb{e233x} that turn the equation for
$g$ and $\omega_\phi$ into higher order differential equations. Therefore the
form of the solution given in \refb{e3.7new}-\refb{e3.9} still holds, while 
\refb{e3.10} and \refb{e3.11} get
modified. Hence the physical scalars still remain constant, equal to their 
asymptotic values, for these solutions.

\end{document}